\documentclass[12pt]{article}
\usepackage{enumerate}
\usepackage{amsmath,amsthm}
\usepackage{amssymb, latexsym}
\usepackage[mathscr]{euscript}
\usepackage{hyperref}
\usepackage{color}
\usepackage{array}
\usepackage{flafter}
\usepackage{layout}
\usepackage{graphicx}
\usepackage{epstopdf}
\usepackage[ansinew]{inputenc}
\usepackage{xfrac}
\usepackage{algorithm}
\usepackage{authblk}

\providecommand{\keywords}[1]{\textbf{\textit{Index terms---}} #1}

\theoremstyle{plain}
  
  \newtheorem{coro}{Corollary}
  \newtheorem{lema}{Lemma}

\theoremstyle{definition}

\theoremstyle{remark}
  \newtheorem{remark}{Remark}

\author[1]{Daniel Andr\'es D\'{\i}az--Pach\'on \thanks{\texttt{Ddiaz3@miami.edu}}}
\author[2]{Robert J. Marks II \thanks{\texttt{ Robert Marks@Baylor.edu}}}
\affil[1]{Division of Biostatistics - University of Miami, Miami, FL}
\affil[2]{Department of Engineering - Baylor University, Waco, TX}

\title{Active information requirements for fixation on the Wright-Fisher model of population genetics}
\date{\today}

\begin{document}

\maketitle

\begin{abstract}
In the context of population genetics, active information can be extended to measure the change of information of a given event (e.g., fixation of an allele) from a neutral model in which only genetic drift is taken into account to a non-neutral model that includes other sources of frequency variation (e.g., selection and mutation).  In this paper we illustrate active information in population genetics through the Wright-Fisher model.
\end{abstract}

\keywords{Active information, coalescence, maximum entropy, Wright-Fisher model.}

\section{Introduction}

Stochastic processes have historically been closely related to the study of biological populations. In fact, biological applications have been central to the development of concepts that are today at the core of probability and statistics. At the beginning of the twentieth century, in a quest to mathematize Darwin's model, Ronald Fisher developed fundamental concepts of statistical inference, and also the basic stochastic process in the analysis of population genetics \cite{Fisher1922}. Fisher has been crowned ``the greatest of Darwin's successors'' \cite{Edwards2011}. His theory, together with Sewall Wright's contributions, served to develop what was subsequently dubbed the Wright-Fisher model and is the focus of this paper.

Interestingly, as central as information has been to genetics since the discovery of DNA and protein synthesis, information theory has been almost universally neglected in the stochastic theory of population genetics. We introduce it here from the perspective of active information under maximum entropy \cite{DiazMarks2020}. Our main goal is to measure the information change in an event when we jump from the \textit{neutral} Wright-Fisher model, in which the only ``force'' in operation is genetic drift, to the \textit{non-neutral} Wright-Fisher model with mutations and selection. In some instances this active information is positive and large, making selection and mutation not free lunches in the sense of Wolpert \& MacReady \cite{Wolpert1997}.

In general, active information is a measure of the degree to which a process deviates from equilibrium. In particular, when it is positive, it measures the amount of guiding information needed to achieve success in certain stochastic searches \cite{Dembski2009,EwertDembskiMarks2012,MarksDembskiEwert2017}. Active information has been applied to analysis of software and modeling attempts to simulate evolution including programs named AVIDA \cite{LenskiEtAl2003}, EV \cite{Schneider2000} and Metabiology \cite{Chaitin2013}. Each model works only because external information has been applied to guide the program to success.  In subsequent studies critiquing of these models, significant active information has been shown to be required for the success of AVIDA \cite{EwertDembskiMarks2009}, EV \cite{MontanezEtAl2010}, and Metabiology \cite{EwertDembskiMarks2013}.

To our knowledge, this paper is the first application of active information to population genetics and the first application of generalized active information \cite{DiazMarks2020} anywhere. Our analysis opens the door to further research on more sophisticated models of population genetics.

\section{The Wright-Fisher model}

In the simplest scenario, the neutral {\em Wright-Fisher model} (see, e.g., \cite{Durrett2008, Etheridge2011}), there are $N$ individuals of a haploid population (i.e., each individual has a single copy of each gene), and each gene has two types of alleles: $A$ and $a$. All individuals of generation $n$ are replaced in the following generation $n+1$ according to a sampling with replacement among all individuals in generation $n$. In other words, if there are $i$ individuals with the $A$ allele in generation $n$, then the probability of having $j$ individuals with that same allele in generation $n+1$ is binomial with parameters $N$ and $i/N$:
\begin{align}\label{WF}
	p_{n, n+1} (i, j) = \binom{N}{j}\left(\frac{i}{N}\right)^j\left(1-\frac{i}{N}\right)^{N-j}.
\end{align}
This type of model is sometimes called the {\em forward model} because it is focuses on future offspring.

The behavior of future generations depends on the initial distribution.  In the absence of any additional knowledge, we assume the $N$ elements are chosen to be either of type $A$ or type $a$ with equal probability, so that the null probability of $j$ alleles being of type $A$ in the first generation is, according to the principle of maximum entropy and assuming that both alleles $a$ and $A$ are present at time 0,
\begin{align}\label{equiWF}
	p_0(i)=\frac{1}{N-1},
\end{align}
for $i= \{1, \ldots, N-1\}$. In fact, it can be shown that the number of $A$ alleles has this distribution, conditionally on the event that both alleles $a$ and $A$ have coexisted for a long time, so that none of them has been lost (see, e.g., \cite[Chapter~8.4]{CrowKimura1970}).
 
 The only ``force'' in the neutral Wright-Fisher model with offspring probability given by (\ref{WF}) is genetic drift; i.e., the changes in alleles proportion from generation to generation are due to genetic drift.  The incorporation of selection and mutation (i.e., a purely Darwinian process) is not imposed in (\ref{WF}). Following Etheridge \cite{Etheridge2011}, we will introduce these two by steps. Notice from (\ref{WF}) that the neutral probability of sampling $A$ is $i/N$, and the neutral probability of sampling $a$ is $1 - i/N$. When we introduce a selection coefficient $s$ these probabilties are modified as follows:
\begin{align}\label{WFSel}
	\begin{array}{ccc}
		\textbf P[A \text{ is sampled}] = \frac{i(1+s)}{i(1+s) + N - i} & \text{ and } & \textbf P[a \text{ is sampled}] = \frac{N-i}{i(1+s) + N - i}.
	\end{array}
\end{align}
Thus, with respect to (\ref{WF}), when $s$ is positive, $A$ is favored in the sampling, in whose case $A$ is said to be \textit{beneficial}; on the other hand, when $s$ is negative, $A$ is not favored in the sampling, in whose case $A$ is said to be \textit{deleterious}. When $s=0$, we go back to the original drift-only scenario.

Now, suppose that every individual of type $A$ mutates to $a$ with probability $\mu_1$, and every individual of type $a$ mutates to $A$ with probability $\mu_2$. In this case, after adding both selection and mutation, the expected proportion of type $A$ offspring in the next generation is given by
        \begin{align}\label{WFSelMut}
        		\theta_i=\frac{i(1+s)(1-\mu_1)}{i(1+s)+N-i} + \frac{(N-i)\mu_2}{i(1+s) + N-i}.
        \end{align}
Notice that the first term at the RHS of (\ref{WFSelMut}) corresponds to the probability of sampling $A$ in (\ref{WFSel}) multiplied by $(1-\mu_1)$; that is,  it considers the type-$A$ individuals that did not mutate to $a$. Analogously, the second term corresponds to the probability of sampling $a$ multiplied by $\mu_2$; that is, it considers the proportion of type-$a$ individuals that mutated to $A$. The two terms added in (\ref{WFSelMut}) give the proportion of type-$A$ individuals in the next generation.

We know from (\ref{WF}) that, given $i$ individuals of type $A$ at time $n$, the conditional distribution of type $A$ individuals at time $n+1$ is  Bin$(N,i/N)$. The addition of selection and mutation changes the conditional distribution to Bin$(N,\theta_i)$, where $\theta_i$ is as in (\ref{WFSelMut}). This is the so-called  {\em Wright-Fisher model with selection and mutation}.

\subsection{Active information in the Wright-Fisher model}

For the Wright-Fisher model, let $\psi \sim$Bin$(N, i/N)$ and $\varphi \sim$Bin$(N, \theta_i)$ be the r.v.'s obtained for the neutral and the non-neutral models described above, respectively.\footnote{We here have reversed the order of notation with respect to \cite{DiazMarks2020}.}  Let the target $T$ be a given event under observation in the space $\{0,\ldots, N\}$. Conditioned on having $i$ type-$A$ alleles in the present, the {\em endogenous information}, $I_\Omega$, is the Shannon information associated with the probability $\psi$ of reaching a target $T$ under an assumption of neutrality.

When neutrality is disregarded, the probability of reaching $T$ changes and the Shannon information it generates is called {\em exogenous information}, $I_1$. The amount by which this information has changed from neutrality to non-neutrality,
            \begin{equation} \label{ChristIsRizen0421} 
            	I_+ = I_\Omega - I_1,
            \end{equation}
is the \textit{active information}. It can be positive, negative or zero. Active information will be negative when the non-neutral model performs worse than the neutral one, in terms of finding the target $T$. %For domains bounded on both sides, maximum entropy is achieved by a uniform distribution. Single sided domains achieve maximum entropy with exponential and geometric distributions while unbounded domains require a normal or gaussian distribution \cite{DiazMarks2020}.

\begin{lema}\label{AIWF}
	\begin{itemize}
	\item[\textit{i.)}] The active information of the success event of drawing one single $A$ allele (with probability $i/N$) of the Wright-Fisher model with selection and mutation, referenced to the neutral model, is given by $I_+ = \log \frac{\theta_i}{i/N}$.
	\item[\textit{ii.)}] In particular, when we add only selection but no mutations, the active information becomes
	\begin{align}\label{aiwf}
		I_+= \log \frac{N + Ns}{N + is}.
	\end{align}
	\end{itemize}
\end{lema}

\begin{remark} This computation of active information in Lemma \ref{AIWF} refers to one child drawing an allele from a pool of $i$ $A$-alleles and $N - i$ $a$-alleles. This corresponds to $\Omega = \{a, A\}$ and $T = \{A\}$. In this case $\psi(A) = i/N$ is not uniform, but it can still be motivated as a `maxent procedure', since the child draws the parent according to a uniform distribution from a pool of $N$ parents.
\end{remark}

\begin{proof}
	\textit{i.} Note that the endogenous information of the success event is given by $I_\Omega = -\log(i/N)$, and the exogenous information of that same event is            $I_1 = -\log \theta_i$. To obtain the active information just apply \eqref{ChristIsRizen0421}.

    \textit{ii.} Replace the actual values of (\ref{WFSel}) in part \textit{i}:
	\begin{align*}
		I_+& = \log \frac{\frac{i(1 + s)}{i(1 + s) + N - i}}{\frac{i}{N}}\\
			&= \log \frac{\frac{i(1 + s)}{N + is}}{\frac{i}{N}}\\
			&= \log \frac{N + Ns}{N + is}.
	\end{align*}
\end{proof}

For $N > i$, from part \textit{ii.} of Lemma \ref{AIWF}, we see that the active information of success is positive whenever $s$ is positive, and it is negative whenever $s$ is negative. It is zero when $s = 0$.

\begin{lema}
	Let $\varphi\sim$Bin$(N, \theta_i)$ and $\psi\sim$Bin$(N, i/N)$. Let  also $T$ be the event that, conditioned on having $i$ type $A$ alleles in the present, we obtain $j$ type $A$ 		alleles in the next generation. Then
	\begin{align*}
		I_+ (\varphi | \psi)(T) =  j \log \left(\frac{\theta_i}{\frac{i}{N}}\right) + (N-j) \log \left(\frac{1-\theta_i}{1-\frac{i}{N}}\right).
	\end{align*}
	In particular, when $\mu_1 = \mu_2 = 0$, the active information becomes
	\begin{align*}
		I_+ (\varphi | \psi)(T) =  j \log(1 + s) + N \log \left(\frac{N}{N + is}\right).
	\end{align*}
\end{lema}

\begin{proof}
	\begin{align*}
		I_+ (\varphi | \psi)(T) &= \log\left(\frac{\theta_i^j\left(1-\theta_i\right)^{N-j}}{\left(\frac{i}{N}\right)^j\left(1-\frac{i}{N}\right)^{N-j}}\right) \\
			&= \log \left(\left(\frac{\theta_i}{\frac{i}{N}}\right)^j \left(\frac{1-\theta_i}{1-\frac{i}{N}}\right)^{N-j}\right) \\
			&= j \log \left( \frac{\theta_i}{\frac{i}{N}} \right) + (N-j) \log \left( \frac{1-\theta_i}{1-\frac{i}{N}} \right).
	\end{align*}
To see the particular case, we replace $\theta_i$ by the probability of sampling $A$ in (\ref{WFSel}) to obtain:
	\begin{align*}
		I_+ (\varphi | \psi)(T) &= j \log \left( \frac{\frac{i(1 + s)}{N + is}}{\frac{i}{N}}\right) + (N-j) \log \left( \frac{\frac{N - i}{N + is}}{\frac{N-i}{N}}\right)\\
			&= j\log \left( \frac{N(1 + s)}{N + is} \right) + (N - j) \log \left( \frac{N}{N + is}\right)\\
			&= j \log(1 + s) + N \log \left(\frac{N}{N + is}\right).
	\end{align*}
\end{proof}
Conveniently, it is easier to calculate the active information of the event $T$ than the individual probabilities $\varphi(T)$ and $\phi(T)$, since the combinations $\binom{N}{j}$ cancel out in the argument of the logarithm.

\begin{coro} 
	When mutations are absent and selection is present, the active information of fixation of the type $A$-allele in generation $n+1$, given that there are $i$ type $A$-alleles in 			generation $n$ (i.e., that $j = N$ in generation $n+1$, given that the proportion of $A$ alleles in generation $n$ is $i/N$), is 
	\begin{align}\label{fixa}
		I_+ = N \log \left( \frac{N + Ns}{N + is} \right).
	\end{align}
\end{coro}

\begin{remark}
	Notice that the active information in (\ref{fixa}) is $N$ times the active information in (\ref{aiwf}). 
\end{remark}

This is a fixation probability of $A$ in a single step, which is highly unlikely unless $i \approx N$ in generation $n$. A more interesting result has to do with the \textit{eventual} fixation of $A$, but the calculation of this probability, particularly for the model with selection and mutation, is extremely complicated. In this case, the  best strategy is to approximate the process with a diffusion, as we do in the next subsection.

\subsection{Active information for the Wright-Fisher model in the limit}

The neutral Wright-Fisher model is a Markov chain with state space given by $\{0, 1, \ldots, N\}$ and transition probabilities as in (\ref{WF}). The states 0 and $N$ are \textit{absorbing}, meaning that once the chain enters in one of these two states, it cannot leave. When the chain enters the state $N$, it means that $A$ has become fixed. When the chain enters the state 0, $A$ goes extinct and $a$ is fixed. The conditional probability of fixation of $A$, given that there are $i$ alleles of type $A$ at the present, is $i/N$ \cite{Durrett2008}. However, this fixation probability and other calculations are extremely complicated when selection and mutations are present and the distribution is as in $\varphi$. For this reason, it is customary to look for diffusion approximations.

A one-dimensional diffusion is a strong Markov process on $\mathbb R$ with continuous paths (see e.g. \cite{Klebaner2012}).  A diffusion $\{X_t\}_{t \geq 0}$ can be expressed as the solution of a stochastic differential equation driven by a Brownian motion with appropriate boundary conditions:
\begin{align}\label{diff}
	dX_t = \mu(X_t)dt + \sigma(X_t)dB_t,
\end{align}
where $B_t$ is a Brownian motion, and $\mu(x)$ and $\sigma^2(x)$ are called the infinitesimal drift and variance of the diffusion. The limiting diffusion of the Wright-Fisher model with selection and mutation is given by (see \cite[Lemma~5.5]{Etheridge2011}):
\begin{align}
	\mu(p) &= \alpha p(1-p) - v_1p + v_2(1-p)\\
	\sigma^2(p) &= p(1-p).
\end{align}
For this limiting diffusion, time has been rescaled to units of $N$ generations: $\alpha = Ns$, $v_1 = N\mu_1$, $v_2 = N\mu_2$, and $p$ is the proportion of type $A$ individuals in the population. In the absence of mutations ($v_1 = v_2 = 0$), the conditional probability of fixation is given by Etheridge (\cite{Etheridge2011}, p. 68):
\begin{lema}\label{fix}
	Suppose that there is no mutation ($v_1 = v_2 = 0$). If the initial proportion of $A$-alleles is $p_0$, the probability $p_{fix}(p_0)$ that the $A$-allele eventually fixes in the 			population (that is the diffusion is absorbed in $p=1$) is
	\begin{align}
		p_{fix}(p_0) =
		\begin{cases}
			\frac{1 - \exp(-2\alpha p_0)}{1 - \exp(-2\alpha)} & \text{ if } \alpha \neq 0,\\
			p_0	&	\text{ if } \alpha = 0.
		\end{cases}
	\end{align}
\end{lema}
From Lemma \ref{fix}, the next Corollary follows directly:
\begin{coro}\label{AIfix}
	Under the same conditions as before, let $T$ be the event that, given that there is an initial proportion $p_0$ of $A$-alleles, the conditional event that the $A$-allele gets fixed is
	\begin{align}
		I_+(T) = \log \frac{\frac{1 - \exp(-2\alpha p_0)}{1 - \exp(-2\alpha)}}{p_0}.
	\end{align}
\end{coro}

Notice that at fixation the search space becomes $\Omega = \{0,1\}$, and the target is $T=\{1\}$.

Of course, once we add mutation, fixation loses all meaning. With mutations, there are no absorbing states in the Markov chain. If a selected allele arises through mutation in an otherwise neutral population, then its actual frequency is $1/N$, so with a little abuse of notation, still calling $p_{fix}(\cdot)$ the first time that the mutated allele has frequency 1, we obtain
\begin{align*}
	p_{fix}\left(\frac{1}{N}\right) = \frac{1 - e^{-2s}}{1 - e^{-2Ns}}.
\end{align*}
In this scenario, Etheridge \cite{Etheridge2011} considers three interesting cases:
\begin{enumerate}
	\item Deleterious alleles: $s < 0$. If $|s| \ll 1$, and $N|s| \gg 1$, then $p_{fix}(1/N) \approx 2|s|e^{-2N|s|}$. The fixation probability of a deleterious allele is exponentially small and it 		decreases with increasing population size.
	\item Beneficial alleles: $s > 0$. If $s \ll 1$, $Ns \gg 1$, then $p_{fix}(1/N) \approx 2s$, almost independent of population size.
	\item Nearly neutral alleles: If $N|s| \ll 1$, then $A$ is nearly neutral and $p_{fix}(1/N) \approx 1/N$.
\end{enumerate}
For these three cases, the active information with respect to the neutral model is given respectively by
\begin{enumerate}
	\item Deleterious alleles: $I_+ \approx \log \left(2N|s| e^{-2N|s|}\right)$. If we measure the information in nats, it becomes $I_+ = -2N|s| +\ln(2N|s|) < 0$.
	\item Beneficial alleles: $I_+ \approx \log 2Ns > 0$. In spite of the probability of fixation being almost independent of population size, active information \textit{is} dependent on 				population size.
	\item Nearly neutral alleles: $I_+ \approx 0$.
\end{enumerate}
Thus, although most alleles (beneficial or deleterious) are lost, ``fitness differences that are too small to be measured in the laboratory ($|s| < 1$) can still play an important role in evolution (if $N|s| \gg 1$)'' \cite{Etheridge2011}. This important role is made explicit by the active information measure. In fact, except on the case of nearly neutral alleles, the population size plays a very important role: for the deleterious case, it makes the active information to decrease linearly in $N$; and for the beneficial allele, it makes the active information to increase logarithmically in $N$.

\section{Active information over nonnegative reals: Coalescence}

The geometric distribution with mean $\mu$ possesses maximum entropy among all distributions over the nonnegative integers with specified mean $\mu$. Letting $\tau\sim\text{Geom}(1/\mu)$ and calling $\psi_\mu$ its distribution, this means that, if we have a search space $\Omega = \{1,2,\cdots \}$ and our only knowledge is that we are looking at a target $T \subset \Omega$ according to a distribution with mean $\mu$, every representation of the search of $T$ must start with the null (endogenous) information in terms of $\tau$, it is the equilibrium from which the active information will be measured (see Table 1 of \cite{DiazMarks2020}).

This sets the stage to think  {\em backwards} of the Wright-Fisher model as the genealogy of a population instead of its offspring. Assume that our prior knowledge is that the population has size $N$. Then, since each new generation is obtained after a sampling with replacement, the probability of two given individuals at the present generation sharing the same father is $1/N$.

Defining $T := \{\tau = k\}$, for $k \in \mathbb Z^+$, we obtain that the probability of a common ancestor $k$ generations before is given by a geometric distribution with mean $N$:
\begin{align}\label{Geo}
	\psi(T) = \left(1-\frac{1}{N}\right)^{k-1}\frac{1}{N},
\end{align}
and the endogenous information is $I_\Omega = -\log \psi(T)$. Suppose that after further analysis we obtain knowledge that the population is actually of size $\nu \neq N$.  This defines an alternative search that uses $-\log \varphi(T)$ bits of information, where $\varphi$ is the mass density of a geometrically distributed  r.v. with mean $\nu$. The active information measured in nats becomes
\begin{align*}
	I_+(\varphi | \psi)(T) &= \log \frac{\left(1-\frac{1}{\nu}\right)^{k-1}\frac{1}{\nu}}{\left(1-\frac{1}{N}\right)^{k-1}\frac{1}{N}}\\
	&= \log\left\{\left(\frac{1-\frac{1}{\nu}}{1-\frac{1}{N}}\right)^{k-1}\frac{N}{\nu}\right\}\\
	&=  \log\left\{\left(\frac{1-\frac{1}{\nu}}{1-\frac{1}{N}}\right)^k \frac{N-1}{\nu -1}\right\}
\end{align*}
If we rescale $k = dN$ and take $\nu$ to be $\mathcal O (N)$, then there is $0 < c < \infty$ such that $\nu \approx cN$ when $N$ is large, and the active information becomes
\begin{align}\label{ComparingGeoms}
I_+(\varphi | \psi) \approx \left(1- \frac{1}{c}\right)d-\log c,
\end{align}
which is a linear function of $d$. From this we notice that the limiting active information is positive for large $d$ when $\nu > N$, i.e. $c > 1$. On the contrary, the limiting active information is  negative (positive) for small $d$ when $\nu > N$ ($\nu < N$), i.e. $c>1$ ($c<1$). This corresponds to our intuition. For instance, if tilting increases the size of the population $(c > 1)$, then we expect longer coalescence times, and this make it easier (more difficult) to find targets that correspond to unusually large (small) coalescence times.

A natural extension of this model is to consider a geometric distribution with success probability $\lambda = 1/N$, which, as $N$ grows, approaches an exponential distribution with intensity $\lambda$. The exponential distribution is also the maxent distribution over the set of all distributions with support on the nonnegative reals and intensity $\lambda$ (see Table 1 of \cite{DiazMarks2020}). Then, when our only knowledge of a search is that it is done on $\{0\} \cup \mathbb R^+$ and that it has a finite mean $\mu = \lambda^{-1} = N$, \cite{DiazMarks2020} tells us that our endogenous search must be guided by an exponential r.v. with mean $\mu$.

Consider the Wright-Fisher model and assume we are taking $N$ generations as our unit of time (i.e., just like 1 minute has 60 seconds, one unit of time here has $N$ generations), then, going backwards, the time to go from $i$ lineages to $i-1$ is exponential with intensity $i(i-1)/2$ and mean $2/[i(i-1)]$. (J. F. C. Kingman was the first to develop the coalescent in \cite{Kingman1982}; Berestycki has an excellent material explaining where the current research is \cite{Berestycki2009}.)

Any change in the distribution, as usual, is contributing information for faster or slower coalescence. To see this,  define the event $T=$ ``The time to coalescence from $i$ to $i-1$ lineages is more than $t$.'' Then any other distribution altering the target, say another exponential with different mean $\mu$, is adding
\begin{align*}
	I_+(T) &= \log\frac{\exp\left\{-\frac{1}{\mu} t\right\}}{\exp\left\{-\frac{i(i-1)t}{2}\right\}}\\
	& = \frac{i(i-1)t}{2} - \frac{1}{\mu} t\\
	& = t \left( \frac{i(i-1)}{2} -\frac{1}{\mu} \right)
\end{align*}
nats of information. Thus, in the same lines of the discrete situation, when $\mu = \frac{2}{i(i-1)}c$, which corresponds to the rescaled coalescence time in units of $N$ for a population of size $cN$, the active information becomes
\begin{align*}
	I_+(T) = \frac{i(i-1)t}{2}\left(1 - \frac{1}{c}\right).
\end{align*}

 Thus, the active information is positive when $\mu > \frac{2}{i(i-1)}$ and negative when $\mu < \frac{2}{i(i-1)}$. As expected, as long as we are considering another exponential distribution, shrinking the mean of the exogeneous search will lessen the probability that the coalescence time is large, whereas augmenting the former will increase the probability of the latter.

\section{Conclusion}

Generalized active information is well-suited to measure the amount of information introduced in a population genetics model when neutral models are replaced by non-neutral ones. In this paper, we focused on the Wright-Fisher model, studying how much information is added by the Darwinian paradigm (considering selection and mutations) with respect to a neutral model that only takes into account genetic drift. Other variation sources (e.g., recombination) can be considered in order to determine the amount of information they introduce to the analysis.

The Wright-Fisher model is the basic introductory model of population genetics. As such, this article opens up a research path in which the neutral and non-neutral versions of more sophisticated models can be compared through active information.

In the pre-limiting model, as it was made explicit in Remark 2, when comparing the Wright-Fisher model with mutation to the neutral Wright-Fisher model, the active information of fixation is $N$ times the active information of the success event of the binomial distributions under consideration.

When we go to the limit, as we saw at the end of Section 2, active information makes explicit fitness differences that cannot be observed in the laboratory but that still are significant asymptotically. In fact, even though the probability of fixation of a beneficial allele in a selection-only model is independent of the population size, active information shows that the information that the selection coefficient introduces plays a very important role as the population size increases. When a deleterious allele is introduced, the population size makes the active information negative; and when a  beneficial allele is introduced, it makes the active information positive.

In summary, we see at least two things: First, selection does not act in the model as an innocuous force. Selection \emph{adds} information. Second, since selection is an information source (and mutations when they are present), it is not a free lunch, and the information it adds is compounded by the population size. Our research stands alongside the work of Basener and Sanford \cite{Basener2018} who through alternative analysis have demonstrated the ineffectiveness of the Wright-Fisher model to create information ex-nihilo.

When active information was originally introduced, its purpose was to measure the amount of information added by a programmer in an alternative search, referenced to a blind one. In this kind of problem the (search) space is always compact. However, non-compact spaces require a generalization of the baseline distribution beyond uniformity in order to be able to account for equilibrium in these spaces, particularly when the space is at least countably infinite. This generalization was developed in \cite{DiazMarks2020}. As such, to our knowledge, this application to population genetics is the first to use generalized active information.

\section{Acknowledgements}

The authors thank the reviewers for the careful reading of the text and their comments. This paper is better because of their suggestions and corrections.

\end{document}